\documentclass[prb,showpacs,superscriptaddress,twocolumn,amsmath,amssymb,amsfonts]{revtex4-1}
\usepackage{amsmath}
\usepackage{bm}
\usepackage{epsfig}
\usepackage{graphics}
\usepackage{dsfont}

\def\beq{\begin{equation}}
\def\eeq{\end{equation}}
\def\beqn{\begin{eqnarray}}
\def\eeqn{\end{eqnarray}}

\renewcommand{\bf}{\mathbf}

\newcommand{\Rmnum}[1]{\expandafter\@slowromancap\romannumeral #1@}

\begin{document}
\title{Gapless edge states of $BF$ field theory and translation-symmetric Z2 spin liquids}
\author{Gil Young Cho}
\affiliation{Department of Physics, University of California,
Berkeley, CA 94720}
\author{Yuan-Ming Lu}
\affiliation{Department of Physics, University of California,
Berkeley, CA 94720}\affiliation{Materials Sciences Division,
Lawrence Berkeley National Laboratory, Berkeley, CA 94720}
\author{Joel E. Moore}
\affiliation{Department of Physics, University of California,
Berkeley, CA 94720} \affiliation{Materials Sciences Division,
Lawrence Berkeley National Laboratory, Berkeley, CA 94720}
\date{\today}


%

\begin{abstract}
We study possible gapless edge states of translation-symmetric Z2 spin liquids. The gapless edge states emerge from dangling Majorana fermions at the boundary. We construct a series of mean-field Hamiltonians of Z2 spin liquids on the square lattice; these models can be obtained by generalization of Wen's exactly
solvable plaquette model. We also study the details of the edge theory of these Z2 spin liquids and find their effective BF theory descriptions. The effective BF theories are shown to describe the crystal momenta of the ground states and their degeneracies and to predict the edge theories of these Z2 spin liquids. As a byproduct, we obtained a way to classify the BF theories reflecting the lattice symmetries.  We discuss in closing three-dimensional Z2 spin liquids with gapless surface states on the cubic lattice.
\end{abstract}


\maketitle

\section{introduction}
The recent discovery of a classification\cite{Schnyder2008,Kitaev2009,Wen2012,Bernard2012} for various topological insulators and superconductors\cite{Hasan2010,Qi2011,Hasan2011} relies on discrete symmetries of the non-interacting fermions. For example, band insulators respecting time-reversal symmetry can be classified in two and three spatial dimensions. In both cases there are two distinct classes of band insulators which cannot be smoothly connected to each other.  For the non-trivial phase of time-reversal symmetric insulators in three dimensions\cite{Fu2007,Moore2007,Roy2009}, gapless states of Dirac fermions emerge on its surface. The gaplessness is protected as long as the time-reversal symmetry is respected at the boundary and the bulk gap remains finite.

This is very different from the physics of the fractional quantum Hall effect (FQHE)\cite{Tsui1982,Laughlin1983}, which is one of the most well-understood and rich examples of topological order\cite{Wen1990b,Wen1995}. On the edge of a fractional quantum Hall state, there is a gapless chiral Luttinger liquid\cite{Halperin1982,Wen1990} which can never be gapped out simply because the net chirality of the bulk state prevents some modes from disappearing.  There is no reference to symmetry needed to explain the stability of the edge states in FQHE. Moreover, the topological order of the FQHE is intimately connected to the gapless edge state\cite{Wen1992b}. The connection between the topological bulk theory and the edge theory is easily understood from effective Chern-Simon theory\cite{Wen1992b,Wen1995} of the FQHE. In this effective theory description, the edge degree of freedom is encoded in the gauge invariance of the Chern-Simons theory with an open boundary.

However, Z2 spin liquids, which are also one of the well-established examples of topological order, do not have gapless edge states in general. The effective $BF$ theory\cite{Hansson2004,Kou2008,Xu2009} description of Z2 spin liquid {\it does} have an edge degree of freedom as in the Chern-Simons theory: the edge theory is a pair of chiral fermions propagating in opposite directions. In the absence of any symmetry, these two fermion modes can backscatter to open up a gap\cite{Hansson2004,Kou2008} in contrast to the ``chiral" Chern-Simons theory of FQHEs. This is true as far as no symmetry is imposed and can be generalized to any `doubled' theory\cite{Freedman2004}. However, it is known that the edge theory of the Abelian `doubled' theory can be gapless\cite{Levin2009,Cho2011} when $U(1)$-charge conservation and time-reversal symmetry are present. The physical example of this is the fractional (and integer) quantum spin Hall effect\cite{Bernevig2006,Levin2009,Levin2011,Neupert2011a,Lu2012,Levin2012a}. As an Abelian `doubled' theory can be re-written formally as the $BF$ theory (at the level of the Lagrangian) and a Z2 spin liquid is described by $BF$ theory, one might think that the edge theory of Z2 spin liquids can be also gapless if the time-reversal symmetry is imposed. This is, in fact, incorrect for Z2 spin liquids.

The reason why Z2 spin liquids fail to have a gapless edge state is traced back to the differences in the charge lattice of the compact gauge theory (or equivalently, the differences in the allowed operators of the edge theory due to quantization). We will see that the ${\mathbb Z}_{2}$ conservation of charge and vortices in Z2 spin liquids is crucial. We will expose this `structural' difference of the formally identical theories in the subsequent discussion in this paper. Nonetheless, the understanding of the gapless edge states of the fractional spin Hall effect indicates that the doubled theories can have gapless edge states. Thus, it implies that Z2 spin liquids will have gapless edge states if the correct symmetries are imposed on top of the topological order. In this paper, we will show that translational symmetry can stabilize the gaplessness of Z2 spin liquids in certain cases. We make a direct connection between the microscopic structure of the physics and the effective $BF$ theory to confirm the gaplessness of the edge states.

Another motivation of this paper is to study the effect of the translational symmetry\cite{Kou2009,Chen2011} on the so-called ``\emph{intrinsic topological order}". The intrinsic topological orders in gapped phases are featured by long-range entanglement\cite{Kitaev2006a,Levin2006}, fractional excitations\cite{Arovas1984}, and topological degeneracies\cite{Wen1990b}. Examples of intrinsic topological orders are the fractional quantum Hall effect states\cite{Laughlin1983} and gapped quantum spin liquids\cite{Wen2002}. Intrinsic topological orders are stable against \emph{any} weak perturbations in the presence of no symmetry, to compare with ``\emph{symmetry-protected topological order}". The symmetry-protected topological (SPT) phases\cite{Chen2011,Gu2012,Wen2012} include topological insulators and topological superconductors\cite{Hasan2010,Qi2011,Hasan2011} as its outstanding examples. One SPT phase is a gapped phase with no fractional excitations and no topological degeneracy, which has gapless boundary states protected by symmetry. Most importantly, a gapped SPT phase can be {\it continuously} connected to the trivial phase if the symmetry is broken\cite{Chen2011b,Chen2011,Wen2012,Lu2012a}. A natural question is: what kind of role does symmetry play for intrinsic topological orders? It turns out that different \emph{symmetry enriched topological (SRT) phases}\cite{Chen2011} can emerge from the same intrinsic topological order, such as different classes of fractional topological insulators\cite{Levin2012a} with $U(1)\rtimes Z_2^T$ symmetry. Therefore fractional topological insulators\cite{Levin2009,Maciejko2010,Swingle2011} and other SRT phases exhibit interesting interplay of fractionalization and symmetry. Here, we show that by imposing translational symmetry new physics emerges in Z2 spin liquids such as gapless edge states and the Majorana zero modes at the lattice dislocations.

The rest of the paper is organized as follows. In section \ref{GENERAL DISCUSSION}, we will study the general edge theory of Z2 spin liquids and show that the edge theory is generally gapped when there is no symmetry imposed on top of the $Z_2$ topological order. We also compare the edge theory of Z2 spin liquids to the edge theory of the quantum spin Hall insulator. In section III, we impose translational symmetry on Z2 spin liquids. We reveal the structure of Wen's plaquette model, which has a gapless edge state, and generalize it to construct mean-field Hamiltonian of Z2 spin liquids with gapless edge states. We also study the edge theory in detail, including the condition for the existence of the gapless edge states and its stability. In section IV, we encode the lattice translational symmetry into the $BF$ theory and make a connection between $BF$ theories and the Z2 spin liquids found in section III. We present a generalization to a three-dimensional model in section V, then summarize our results and raise open questions in section VI.

\section{General Discussion of edge theory of Z2 spin liquids}\label{GENERAL DISCUSSION}
In Z2 spin liquids, there are two low-energy excitations, namely the spinon and the vison. Spinons (visons) carry electric charge $e=1$ (magnetic charge $m=1$) in the underlying ${\mathbb Z}_{2}$ gauge theory. They can be described by $BF$ theory with the charge lattice $e \in {\mathbb Z}_{2}$ and $m \in {\mathbb Z}_{2}$. We denote the electric current $J_{\mu}$ and magnetic current $j_{\mu}$ which couple to $a_{\mu}$ and $b_{\mu}$ minimally. Then, the low-energy theory for the $Z_2$ spin liquid\cite{Hansson2004,Kou2008,Xu2009} is
\beq
L = \frac{1}{\pi} \varepsilon^{\mu\nu\lambda} a_{\mu} \partial_{\nu}b_{\lambda} - a_{\mu}J^{\mu} - b_{\mu}j^{\nu} + O(\partial a, \partial b)^{2}.
\label{BF}
\eeq
As noted before, a formally similar $BF$ theory emerges as the effective theory for the spin Hall effect with a Kramer pair of counter-propagating gapless edge modes. To manifest this similarity, we diagonalize the $BF$ Lagrangian into two copies of Chern-Simon theory, i.e., we write $A_{\mu} = a_{\mu}+b_{\mu}$ and $B_{\mu}=a_{\mu}-b_{\mu}$ to obtain
\beq
L = \frac{1}{4\pi} \varepsilon^{\mu\nu\lambda}A_{\mu}\partial_{\nu}A_{\lambda} - \frac{1}{4\pi} \varepsilon^{\mu\nu\lambda}B_{\mu}\partial_{\nu}B_{\lambda} - A \cdot \chi-B \cdot \tau,
\label{DCS}
\eeq
where the dot product ($A\cdot J = A_{\mu}J^{\mu}$) are understood. We introduced the source currents $\chi = \frac{J+j}{2}$ and $\tau = \frac{J-j}{2}$ for $A_{\mu}$ and $B_{\mu}$. The charge lattice associated with $\chi$ and $\tau$ can be deduced from that of $J$ and $j$ (i.e., of $e$ and $m$). We begin with $(e, m=0)$ where $e$ is defined modular $2$. For $e=0$ and $m=0$, $(\chi, \tau) = (0,0)$. For $e=1$ and $m=0$, $(\chi, \tau) = (1/2,1/2)$. For $e=2 \sim 0$ mod $2$ and $m=0$, $(\chi, \tau) = (1,1) \sim (0,0)$. Similar consideration shows that $(\chi, \tau) = (1,-1) \sim (0,0)$. Thus, in the diagonalized doubled Chern-Simon theory, $(\chi, \tau) = (1,\pm1) \sim (0,0)$ is the same as $(e,m) = (2,\pm2) \sim (0,0)$ in $BF$ theory. Now, we construct the edge theory from Eq.\eqref{DCS}.
\beq
L_{edge} = \psi^{\dagger}_{R}(\partial_{t}-v\partial_{x})\psi_{R} + \psi^{\dagger}_{L}(\partial_{t}+v\partial_{x})\psi_{L}
\label{edge}
\eeq
This edge theory, in general, is unstable to opening up a gap\cite{Hansson2004,Kou2008}. This can be easily seen from ${\mathbb Z}_{2}$ conservation of spinons and visons. The ${\mathbb Z}_{2}$ gauge charge conservation allows us to add the mass term $\sim \psi^{\dagger}_{R}\psi_{L}, \psi^{\dagger}_{L}\psi_{R}$ and $\psi_{R}\psi_{L}, \psi_{R}^{\dagger}\psi_{L}^{\dagger}$.  This is because $\psi^{\dagger}_{R}\psi_{L}$ , $\psi^{\dagger}_{L}\psi_{R}$, or $\psi_{R}\psi_{L}, \psi_{R}^{\dagger}\psi_{L}^{\dagger}$ carry charges $(\chi, \tau) = (1,\pm1)$ which are equivalent to no charge $(\chi, \tau) = (0,0)$ in ${\mathbb Z}_{2}$ theory. Hence, we conclude that there is no protected edge state for a general Z2 spin liquid without additional symmetries, independent of microscopic constructions. However, this could be changed when the Z2 spin liquid is supplemented by a symmetry. Additional symmetries on the topological order can restrict the form of the mass terms and stabilize the gapless edge state.

We now show that the time-reversal symmetry cannot stabilize the `gaplessness' of the edge states of Z2 spin liquids (if it does not possess any `strong index'\cite{Fu2007} for the underlying fermionic spinons). We take the edge theory Eq.\eqref{edge} and consider the time-reversal symmetry operation on the $A_{\mu}$ and $B_{\mu}$. Due to the fact that under time reveral $T: A_{\mu} \rightarrow (-1)^{\mu}B_{\mu}$ and $B_{\mu} \rightarrow (-1)^{\mu}A_{\mu}$  (with the definition $(-1)^{\mu} = 1$ for $\mu =0$ and $(-1)^{\mu} = -1$ for $\mu = 1,2$) under the time-reversal operation $T$, the time-reversal operation effectively acts as the exchange of two fermionic fields, i.e., $T: \psi_{R} \rightarrow \psi_{L}$ and $ \psi_{L} \rightarrow e^{i\theta}\psi_{R}$ up to the $U(1)$ phase factor $e^{i\theta}$. We can take $\theta = \pi$ or $0$ to be consistent with the time-reversal operation $T^{2}=(-1)^{\hat{N}_f}$ or $T^{2}=1$, where $\hat{N}_f$ represents the total fermion number operator. This operation should be supplemented with $v \rightarrow -v$. Hence, the kinetic term for the edge theory Eq.\eqref{edge} is time-reversal symmetric. We discuss the two different cases $T^{2}=1$ and $T^{2}=(-1)^{\hat{N}_f}$ independently. First, for $T^{2}=1$ the time-reversal symmetry allows mass terms of the form $\psi_{R}^{\dagger}\psi_{L}$, with an equal amplitude for $\psi_{L}^{\dagger}\psi_{R}$, which are capable of gapping the edge. Other terms such as $\psi_{R}\psi_{L}$ are not allowed. As a whole, the time reversal symmetric edge theory for Z2 spin liquid is
\beq
L = L_{edge} (\psi_{R}, \psi_{L}) + m (\psi_{R}^{\dagger}\psi_{L} + \psi_{L}^{\dagger}\psi_{R}) + h.c.,
\eeq
where $L_{edge}$ is the kinetic term Eq.\eqref{edge}. For $T^{2}=(-1)^{\hat{N}_f}$, $\psi_{R}^{\dagger}\psi_{L}$ and $\psi_{L}^{\dagger}\psi_{R}$ are not allowed. Instead, there are allowed terms such as $\psi_{R}\psi_{L}$, and they are enough to gap out the edge state. Here we assumed that there are no strong indices for the underlying spinons. The strong index includes the ${\mathbb Z}_{2}$-indices for DIII class and ${\mathbb Z}$- indices for C/D classes\cite{Schnyder2008,Kitaev2009} for band structures of the Schwinger fermions\cite{Affleck1988a,Baskaran1987,Baskaran1988,Kotliar1988,Affleck1988b,Wen1996} (fermionic spinons). When the strong index of the time-reversal symmetry is nontrivial, then this phase automatically has the gapless edge state which is either a pair of helical Majorana modes (DIII class) or chiral Majorana modes (C or D classes). However, these phases are not described by $BF$ theory\footnote{If the Schwinger fermion band structure lies in Class C, it corresponds to a chiral spin liquid whose effective theory is a $U(1)$ Chern-Simons theory. In the case of class D and DIII, the corresponding spin liquids host non-Abelian quasiparticle excitations which could in general be described by non-Abelian Chern-Simons theory.} and will not be discussed in this work.

For the quantum spin Hall effect on the other hand, we have $U(1)$ charge conservation and the time-reversal symmetry $T$ with $T^{2}=(-1)^{\hat{N}_f}$. The two conditions exclude\cite{Levin2009} the two mass terms $\psi^{\dagger}_{R}\psi_{L}$, $\psi_{R}\psi_{L}$ which {\it were} allowed in ${\mathbb Z}_{2}$ theory. Hence, the edge excitations of the quantum spin Hall state are gapless, in contrast to that of Z2 spin liquids.

In the remaining sections of this paper, we will consider the effects of the translational symmetries on the states localized at the edges of Z2 spin liquids. We find that translational symmetry can stabilize the gapless modes at the edges of some Z2 spin liquids.

\section{Wen's plaquette model and mean field theory of ${\bf Z}_{2}$ spin liquids}

\begin{figure}
\includegraphics[width=1\columnwidth]{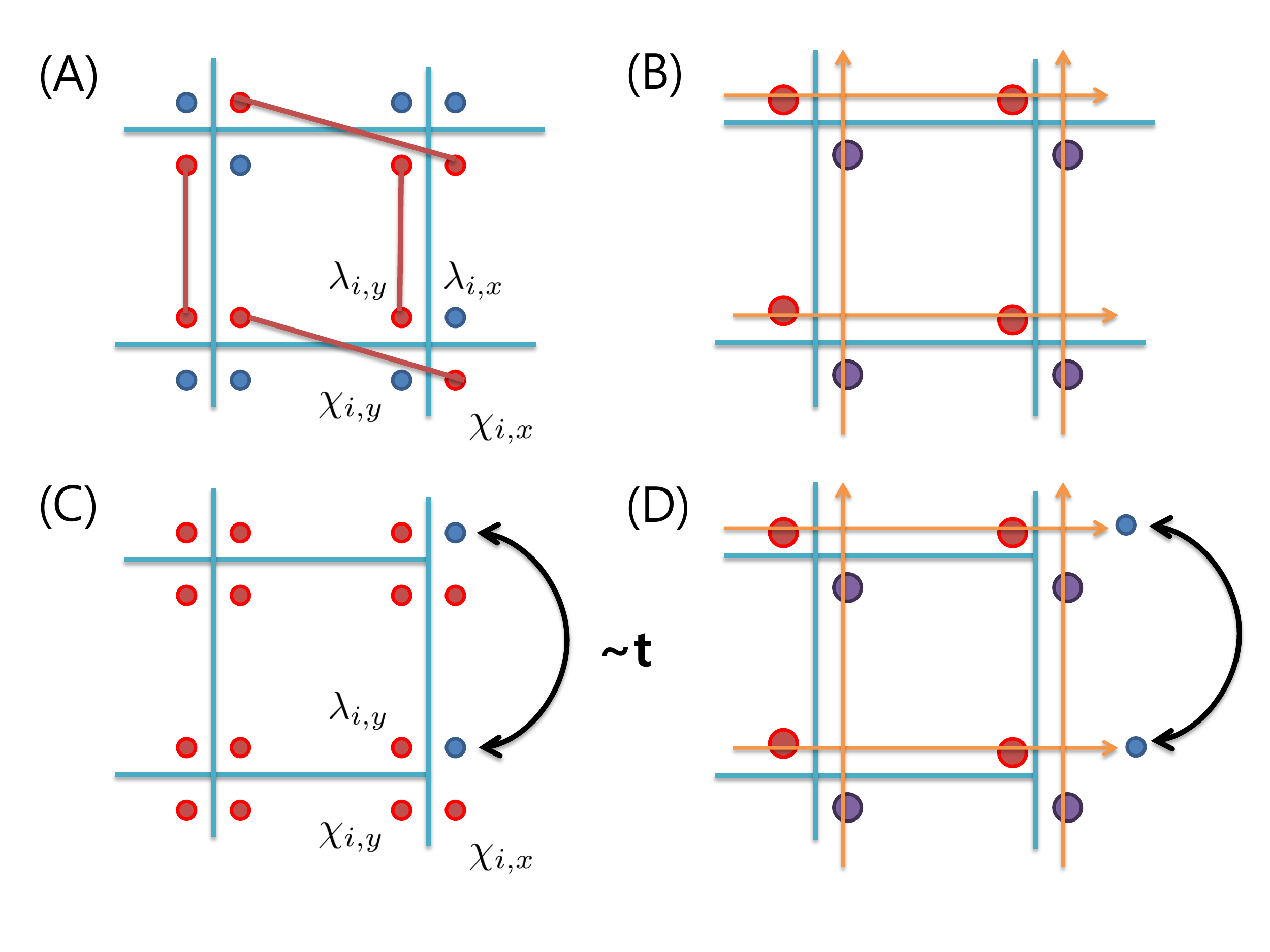}
\caption{Illustration of Wen's plaquette model and its edge state. (A) Plaquette operator $\prod_{<ij>\in P} U_{ij}$. A plaquette operator pairs up four Majorana fermions around the square. Here, blue circles represent four Majorana fermions per site, and red lines connecting two blue circles represent $U_{ij}$ appearing in the plaquette operator. (B) In terms of the complex fermions $f = \lambda \pm i \chi$ in Eq.\eqref{Parton}, Wen's plaquette model can be understood as the stacked one-dimensional Kitaev chains parallel to ${\hat x}$ and ${\hat y}$. (C) Edge states of Wen's model where the edge is along ${\hat y}$ direction. The plaquette operators in Hamiltonian Eq.\eqref{Wen} cover all the Majorana fermions except a {\it single} Majorana fermion per site on the edge. The blue circles represent dangling Majorana fermions while red circles are Majorana fermions covered by the plaquette operators. These blue dangling modes will hop around by the magnitude $\sim t$ when we are away from the exactly solvable point. (D) The dangling Majorana fermions can be equally well understood via the stacked Kitaev chain picture. When the Kitaev chain is cut, the edge always hosts a single Majorana fermion per chain. It is clear that the edge states of (C) and (D) are exactly the same. }
\label{Fig1}
\end{figure}

Wen's plaquette model\cite{Wen2003} is an exactly solvable model on the square lattice where the spin degrees of freedom are fractionalized into Majorana fermions. The model can be formulated as the pure ${\mathbb Z}_{2}$ gauge theory\cite{Kitaev2003}
\beq
H = -g \sum_{P} F_{P} = -g \sum_{P} \prod_{<ij>\in P} U_{ij},
\label{Wen}
\eeq
where $P$ denotes a plaquette of the direct lattice. Here, $U_{ij} = \pm 1$ is the ${\mathbb Z}_{2}$ gauge field on the link $<ij>$, and thus $ \prod_{<ij>\in P} U_{ij} = F_{P}$ is the field strength on the plaquette $P$. A faithful representation of this gauge theory is $U_{i,i+{\hat x}} = i \lambda_{i,x}\chi_{i+{\hat x},x}$, $U_{i,i+{\hat y}} = i \lambda_{i,y}\chi_{i+{\hat y},y}$ with translational symmetry.  There are four Majorana fermions $(\lambda_{i,x},\lambda_{i,y}, \chi_{i,x}, \chi_{i,y})$ per site, and the dimension of the Hilbert space per site is $2^{4/2}= 4$ states per site (see Fig.\ref{Fig1}). To faithfully represent a spin-1/2 system, we need to halve this Hilbert space, i.e., the Hilbert space per site should have dimension $2$. This is done by taking the ${\mathbb Z}_{2}$ redundancy into account: the theory Eq.\eqref{Wen} is invariant under $U_{ij} \rightarrow s_{i}U_{ij}s_{j}$, $s_{i} = \pm1$. The model can be more clearly represented if we introduce two complex fermions
\beq
f_{i,u} = \lambda_{i,x} + i\chi_{i,x}, \quad f^{\dagger}_{i,d} = \lambda_{i,y} + i\chi_{i,y},
\label{Parton}
\eeq
such that the spin ${\vec S}_{i}$ on the site $i$ is represented via $\frac{1}{2}\sum_{\alpha,\beta} f^{\dagger}_{i,\alpha}{\vec \sigma}^{\alpha,\beta}f_{i,\beta}$. Then, the Hamiltonian Eq.\eqref{Wen} reduces to
\beq
H = -g \sum_{P} S_{i,x}S_{i+{\hat x},y}S_{i+{\hat x} + {\hat y}, x}S_{i+\hat{y}, y}.
\eeq
From the model Eq.\eqref{Wen}, we can immediately see the existence of the edge degree of freedom. As the Majorana fermions in bulk will be paired within the plaquette, there will be dangling Majorana fermions\cite{Wen2003} at the boundary because of the edge cuts a plaquette in half (see Fig.\ref{Fig1}). It was already noticed\cite{Wen2003,Kou2008} that this model (with $g>0$) has a flat band of edge states on the boundary along ${\hat x}$- and ${\hat y}$- directional cut (see Fig.\ref{Fig1}). Hence, to look for gapless edge states for Z2 spin liquids protected by translational symmetry, we can consider similar models which supports dangling Majorana fermions on the edge. We will generalize the structure of Wen's model to generate a series of mean-field Hamiltonian for Z2 spin liquids with gapless boundary states in two and three spatial dimensions. We also discuss the stability of the gapless boundary states.

We now look into Wen's model carefully. To illuminate the underlying structure of Wen's model, we study the mean-field Hamiltonian for the fermionic spinons \eqref{Parton}.
\beq
H_{mean} (\{f_{i,u}, f_{i,d}\}) = \sum_{ij} \left[ f^{\dagger}_{i,\sigma} u_{ij}^{\sigma,\tau} f_{i,\tau} +f_{i,\sigma} \eta_{ij}^{\sigma,\tau} f_{i,\tau} \right] + h.c.
\eeq
We notice, by plugging Eq.\eqref{Parton}, that $f_{i,u}$ and $f_{i,d}$ are decoupled completely and form the one-dimensional Majorana fermion chains\cite{Kitaev2001} (in the weak-pairing phase) along ${\hat x}$-  and ${\hat y}$- axis. Explicitly, we now have
\beq
H_{mean} (\{f_{i,u}, f_{i,d}\}) = h_{1D,{\hat x}} (\{f_{i,u}\}) +  h_{1D,{\hat y}} (\{f_{i,d}\}),
\label{Kitaev}
\eeq
where $h_{1D, {\hat n}} (\{f\})$ is the one-dimensional Kitaev model along ${\hat n}$ in the weak pairing regime. This explains why we have dangling Majorana fermions on the boundary along ${\hat x}$- and ${\hat y}$- directional cut of the lattice (See Fig.\ref{Fig1}). We will see that there is a gapless Majorana mode formed by the dangling Majorana fermions as long as there are odd number of Majorana fermions per unit cell on the edge. Before discussing the nature of the gapless edge states, we notice that we can easily generalize Wen's model by deforming one of $h_{1D, {\hat n}}(\{f\})$ Eq.\eqref{Kitaev} into the trivial phase while keeping the constraint `one spinon per site'. Then, this Z2 spin liquid will break the $C_4$ rotational symmetry of square lattice while keeping the translational symmetry. For example, let us consider the case where $h_{1D, {\hat x}}(\{f_{i,u}\})$ is replaced by the trivial $Q(\{f_{i,u}\})$ (this Hamiltonian $Q(\{f\})$ is not necessarily one-dimensional but is fully gapped):
\beq
H_{mean} (\{f_{i,u}, f_{i,d}\}) = Q(\{f_{i,u}\}) +  h_{1D,{\hat y}} (\{f_{i,d}\})
\eeq
has dangling Majorana fermions if the edge is not parallel to ${\hat y}$. Similarly, we can generate a Z2 spin liquid which has dangling Majorana fermions if the edge is not parallel to ${\hat x}$. For Z2A phase\cite{Wen2002,Wen2003} whose spinon band structure preserves translation symmetry explicitly, we can have another possible Z2 spin liquid where we align the Kitaev chain along ${\hat n} = {\hat x} \pm {\hat y}$ (we will see later that this state has the same characteristics to the Wen's model). Hence we have three translational symmetric Z2A spin liquids with the gapless edge states. Though these Z2A spin liquids have the same ${\mathbb Z}_{2}$ topological order and the same translational symmetry, we can distinguish them further by looking into the crystal momenta\cite{Kou2008,Kou2009} of the ground states and degeneracies, and the edge states. In the next section, we use these crystal momenta and degeneracies to find the effective $BF$ theory for the Z2 spin liquids.

\begin{figure}
\includegraphics[width=1\columnwidth]{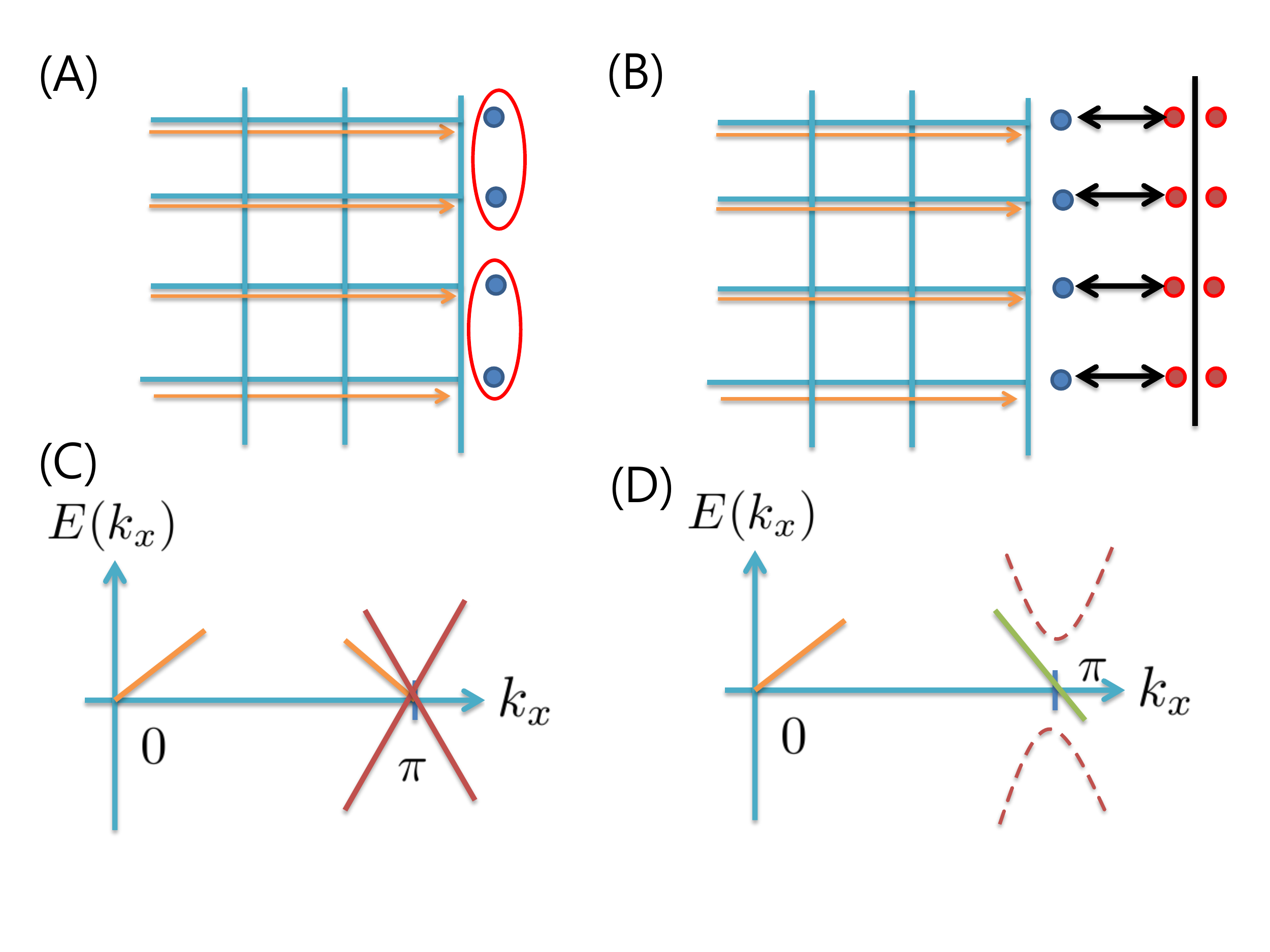}
\caption{Stability of the gapless edge states. (A) To gap out the dangling Majorana fermions, the Majorana fermions should pair up and dimerize. However, it is clear from the figure that the dimerization doubles the unit cell along the boundary. This is similar to the physics of surface states of weak topological insulators~\cite{Mong2012,Ringel2011}. (B) Coupling of the dangling Majorana fermions to the quantum Ising chain at criticality. As the Hilbert space per site of the Ising chain has the dimension $2 = (\sqrt{2})^{2}$, the Ising chain can be thought as the chains of {\it two} Majorana fermions per site. So when the Ising chain is coupled to the dangling Majorana fermions, there are {\it three} Majorana fermions per site which is guaranteed to be stable by counting Eq.\eqref{Counting}. (C) This picture can be further supported by looking at momentum space. The Ising chain at criticality is equivalent to a helical Majorana fermion state at $k = \pi$ and the edge state consist of the dangling Majorana fermions of the spin liquid has one mode at $k=0$ and $k=\pi$. Thus there are total {\it three} modes at $k=\pi$ ($k=0$ is decoupled from $k=\pi$ as the coupling between $k=0$ and $k=\pi$ would double the unit cell). (D) When the interaction between the Ising chain and the edge state of the Z2 spin liquid is turned on, we will be left with one {\it reconstructed} Majorana mode at $k=\pi$ and another mode at $k=0$. }
\label{Fig2}
\end{figure}

Now that we have some conditions for the existence of the edge states, we study their detailed features. We first discuss the relation between the edge degrees of freedom and the bulk gap. We argue that there is a gapless edge state as long as the bulk is gapped and the translational invariance at the edge remains unbroken (See Fig.\ref{Fig2}). To remove the boundary Majorana fermions, we need to pair them up into complex fermions. Due to the translational symmetry, a Majorana fermion on the boundary cannot pair up on the boundary if there is one single Majorana fermion per unit cell. What the Majorana fermion can do, instead of dimerizing on the boundary, is to tunnel through the bulk and pair-annhilate with the Majorana fermion on the opposite boundary. However, the tunneling length\cite{Kitaev2001} is $\sim 1/\sqrt{|\Delta|}$ where $\Delta$ is the bulk gap. Hence, we need to close the gap and go through the bulk phase transition to lose the dangling Majorana fermions at the boundary. In the next section, we will construct the effective $BF$ theory for these phases, and the effective theory should describe a finite range of the phase diagram as it is gapped and the edge state of the topological field theory is stable as long as the gap remains finite.

The above `gaplessness' of the edge states depends on the number of Majorana fermions per unit cell on the boundary. If there are even numbers of the dangling Majorana fermions per unit cell, then Majorana fermions can pair up on the boundary without breaking the translational symmetry along the boundary. The `parity' of the number of the dangling Majorana fermions on the arbitrary boundary can be easily computed,  and there is an odd number\cite{Ran2010} of the dangling Majorana fermions per unit cell on the boundary if
\beq
J_{{\hat n}} = \frac{{\hat n}}{2} \cdot {\vec G} = \pi  \text{ mod } 2\pi,
\label{Counting}
\eeq
where ${\hat n}$ is the vector defined for the Hamiltonian of the Majorana fermion chain $h_{1D, {\hat n}}(\{f\})$, and ${\vec G}$ is the reciprocal vector orthogonal to the boundary. If there is more than one chain, we simply add up all $J_{{\hat n}}$ mod $2\pi$. Then we have gapless edge modes protected by the translational symmetry if the sum is $\pi$ mod $2\pi$. In fact, the same index is used to study the localized states at the dislocations of weak topological insulators/superconductors\cite{Ran2009,Ran2010}. Intuitively, ${\hat n} \cdot {\vec G}$ in Eq.\eqref{Counting} can be thought as the ``flux density'' of Kitaev chains (which are parelle to ${\hat n}$) passing through the area perpendicular to ${\vec G}$ (which is analogous to the definition of the flux through a surface in the elementary calculus, i.e., $F = {\vec E} \cdot {\hat S}$ defines the flux of the vector field ${\vec E}$ passing through the area perpendicular to the vector ${\hat S}$). Due to the `parity' (or even/odd-ness) of the dangling Majorana fermions, we need to take mod by $2\pi$ with the appropriate normalization. To demonstrate this, we work out a few examples applying Eq.\eqref{Counting}.

- {\it example 1.} edge state of Wen's plaquette model with the edge parallel to ${\hat y}$. As the edge is along ${\hat y}$-direction, the reciprocal vector defining the edge is ${\vec G} = (2\pi, 0)$. In Wen's plaquette model, we have the two Kitaev chains where each chain is identified with ${\hat n}_{1} = (1,0)$ and ${\hat n}_{2}=(0,1)$. We immediately have $J_{{\hat n}_{1}} = \pi$ and $J_{{\hat n}_{2}} = 0$ and thus we have a dangling Majorana fermion per site because $J_{{\hat n}_{1}} + J_{{\hat n}_{2}} \text{ mod } 2\pi = \pi$. This agrees with the previous intuitive understanding depicted in Fig.\ref{Fig1}. 

- {\it example 2.} edge state of Wen's plaquette model with the edge parallel to ${\hat x}+{\hat y}$. This edge can be identified with the reciprocal vector ${\vec G} = (2\pi, -2\pi)$. As before, we have the two Kitaev chains where each chain is identified with ${\hat n}_{1} = (1,0)$ and ${\hat n}_{2}=(0,1)$, and thus $J_{{\hat n}_{1}} = \pi$, $J_{{\hat n}_{2}} = -\pi$ and $J_{{\hat n}_{1}} + J_{{\hat n}_{2}} \text{ mod } 2\pi = 0$, i.e., there are even number of Majorana fermions per unit cell at the boundary and hence there is no gapless helical Majorana mode protected by the translational symmetry. Indeed, if we look at the edge defined by ${\vec G} = (2\pi, -2\pi)$, there are two Majorana fermions (per unit cell on the boundary) coming from the Kitaev chains along ${\hat x}$- direction and Kitaev chains along ${\hat y}$- direction on the boundary which immediately implies that there is no stable gapless edge state.

If we have a single Majorana fermion per unit cell, we can study the spectrum of the boundary modes. When the perturbation away from this ideal Hamiltonian is given, the dangling Majorana fermions will hybridize with the nearest neighbors and start to disperse
\beq
H_{edge} = it \sum_{j} \eta_{j}\eta_{j+1},
\eeq
where the position of the boundary Majorana fermion is labeled by the index $j \in {\mathbb Z}$ and $t$ is the effective hopping parameters on the boundary. The spectrum of this Majorana fermion is given by $E(k) = 2t \sin(k)$ for $k \in (0, \pi)$ (this Majorana fermion problem is, in fact, related to the fermion doubling problem). To gap out the spectrum, we need a perturbation with a matrix element connecting $k = 0$ and $k = \pi$. However, this interaction doubles the unit cell on the boundary (similar to the perturbations required to eliminate the surface of the weak topological insulator\cite{Mong2012,Ringel2011}), and hence it is prohibited by the lattice translational symmetry. We note that the right-mover and the left-mover of the gapless edge theory have different center of mass momentum. This feature is also reproduced by the effective $BF$ theory.

From this edge spectrum, the counting of Majorana fermions, and the translational symmetry, we now show that this edge theory mimics the stability of the edge state of a quantum spin Hall insulator against the coupling to ordinary one-dimensional Luttinger liquids. When the edge state of the spin Hall effect interacts with  ordinary Luttinger liquids, the edge states are reconstructing themselves to the edge theory of spin Hall insulators with the renormalized coefficients. For the edge of the Z2 spin liquids, we can couple our edge theory to the quantum Ising chain at the critical point, which is a gapless helical Majorana state (see Fig.\ref{Fig2}). The helical modes (the left-mover and the right-mover) of the quantum Ising chain are located at $k=\pi$. When the helical modes are placed on top of the boundary Majorana modes of Z2 spin liquids, the right-mover at $k=0$ is intact as it cannot interact with the modes at $k=\pi$. At $k=\pi$, there are three modes (two from the critical Ising chain and one from the original QSH edge): one right-mover, two left-movers. The translational-invariant potential will have matrix elements connecting $k=\pi$ to itself, allowing one right-mover and one left-mover to pair up. Hence, we are left with one left-mover at $k=\pi$ which is reconstructed from the helical Majorana modes of the quantum Ising chain and the original left-mover at $k=\pi$. Thus, the boundary remains gapless even if it interacts with the critical Ising chain. It is not difficult to see that other systems, such as helical Dirac edge, cannot gap out the original edge modes as they have even numbers of the helical Majorana fermions. The easiest way to understand this behavior is as follows; the size of the Hilbert space per site in the quantum Ising chain is $2$, and we need two Majorana fermions per site to represent the Hilbert space. This implies that there are three Majorana fermions per unit cell (see Fig.\ref{Fig2}), and thus we expect the spectrum to be gapless after the dimerization of two of them. Hence we can conclude, based on the counting argument and the translational symmetry, that the edge is robust against to coupling to any one-dimensional gapless system.

It is also interesting to note that this edge spectrum implies that we will have a zero energy Majorana state at the lattice dislocation in Z2 spin liquids\cite{Ran2010}. For Wen's plaquette model, the dislocation of the Burger's vector ${\hat x}$, ${\hat y}$ will trap the zero energy Majorana states. On the other hand, we will have the zero energy state of the Burger's vector ${\hat m}$ orthogonal to ${\hat n}$ if Z2 spin liquids contains a Majorana fermion chain along ${\hat n}$ defined above in this section.

The discussion in this section is based on the mean-field Hamiltonian for the Z2 spin liquids. However, the spectrum is fully gapped, and hence the fluctuation over mean-field solutions are supposed to be small. Still, one might think that it is unclear if our result remains the same when both four-fermion interactions and gauge fluctuations are included. These questions, perhaps, cannot be answered in the mean-field theory. Therefore we take a different path to answer the question. In the next section, we will find the effective $BF$ theory for these spin liquids, and the topological $BF$ theory should describe a phase instead of a point in the phase diagram. Hence the $BF$ theory predicts the phase to have gapless edge states on the boundary protected by translational symmetry as long as the gap remains finite.

\section{effective $BF$ theory and its characterization}

In this section, we will construct the effective $BF$ theory of the Z2 spin liquids considered in the previous section. In previous studies\cite{Wen2003,Kou2008} it was shown that the translational symmetry restricts the form of the mass into the specific form so that the edge state becomes gapless. And the same effective theory captures the crystal momenta of the ground states and topological degeneracies in a square lattice with periodic boundary condition. Here we extend previous results in the way that we clarify the spectrum of the edge theory and its connection to the microscopic model discussed in the previous section (in fact, the spectrum of the edge states is crucial for the gaplessness as we saw in the microscopic discussion). Furthermore, we demonstrate the way of classifying $BF$ theory in the presence of lattice symmetries.

The essence of the effective theory is to encode the non-trivial transformations for the gauge fields in $BF$ theory \eqref{BF} under translations (in the Coulomb gauge $a_{0}=b_{0}=0$)
\beq
t_{x,y}: a_{i} \rightarrow b_{i}, \quad b_{i} \rightarrow a_{i},
\eeq
leaving the theory Eq.\eqref{BF} invariant. $t_{x,y}$ denote lattice translations along $x$ and $y$ directions. This directly implies that the edge theory\cite{Kou2008} should be symmetric under $\psi^{\dagger}_{L} \rightarrow \psi_{L}$ and $\psi_{L} \rightarrow \psi^{\dagger}_{L} $ in the edge theory Eq.\eqref{edge}. This symmetry, hence, restricts the form of the masses for the edge theory into
\beq
L_{mass} \sim a \psi^{\dagger}_{R} (\psi^{\dagger}_{L}+ \psi_{L}) + h.c.,
\label{mass}
\eeq
To see the effect of this particular mass term Eq.\eqref{mass}, we introduce $\psi_{R} = \eta_{R}+i\chi_{R}$ and $\psi_{L} = \chi_{L} + i\eta_{L}$ where $\chi_{R/L}, \eta_{R/L}$ are real Majorana fermions. Then, the mass Eq.\eqref{mass} gaps out $\chi_{L}$ and one of two right-moving Majorana modes $\chi_{R}, \eta_{R}$ (e.g. $\chi_R$ if $a$ is a real number), and hence we are left with one right-mover and one left-mover. It is not difficult to see $\eta_{L}$ is located at $k=\pi$ as $\psi_{L}$ transforms to $\psi^{\dagger}_{L}$ under unit lattice translation.

We will follow this lesson to construct and (partially) classify the effective $BF$ theory for the Z2 spin liquids discussed in the previous section. However, the classification in this section is based on the educated guesses constructed in the previous studies. The philosophy of this classification and characterization of $BF$ theory is, thus, phenomenological. Nevertheless, we will find the correct $BF$ theory for the Z2 spin liquids with the gapless edge states. More specifically, we will find that the $BF$ theory matches the the crystal momenta and topological degeneracies, and the nature of the edge theories of Z2 spin liquids. This justifies the correctness of the effective $BF$ theories.

\subsection{Crystal momenta and degeneracies of the Z2 spin liquids}
To find $BF$ theory for the Z2 spin liquids, we first need to compute the crystal momenta and degeneracies for the spin liquids. We assume that the mean-field Hamiltonian is enough for computing these quantities, i.e., the gauge fluctuation and interactions beyond the mean-field states do not change the crystal momenta and degeneracies.

Here, we briefly show how to compute\cite{Kou2008} the crystal momenta and degeneracies of the Z2 spin liquids. The calculation is based on the mean-field theory for the Z2 spin liquids. In the mean-field theory, we can assume that there are two fermions ${\bf \Psi}^{T}= (\psi_{1}, \psi_{2})$ per site. To connect to the spin-1/2 Hamiltonian, we identify $\psi_{1} = f_{\uparrow}$ and $\psi_{2} = f^{\dagger}_{\downarrow}$ where $\uparrow/\downarrow$ denotes the spin up and down in basis of $S_{z}$, i.e., ${\vec S} = \frac{1}{2} \sum_{\alpha,\beta=\uparrow,\downarrow} f^{\dagger}_{\alpha} {\vec \sigma}^{\alpha\beta} f_{\beta}$. This parton construction enlarges the Hilbert spaces $ \{ |\Psi_{mean}>\}$, and we need to project it down to the physical states $\{|\Phi_{spin}>\}$. The projection is given by
\beq
|\Phi_{spin}> = P |\Psi_{mean}>,
\eeq
where $P$ projects the mean field state of fermions into the physical spin state with one fermion per site. In this paper, we specifically concentrate on the case of $Z_2$ spin liquid, where the fluctuating gauge field is $Z_{2}$ gauge field. According to projective symmetry group classification\cite{Wen2002}, there are two types of Z2 spin liquids if we have {\it only} translational symmetry: Z2A spin liquids (so-called zero flux states whose spinon band structure explicitly preserves translational symmetry) and Z2B spin liquids (so-called $\pi$-flux states whose spinon band structure doubles the unit cell). Here we focus on Z2A spin liquids. We introduce the fermion operator for ${\bf \Psi}_{\bf k}$ in the momentum space ${\bf \Psi}_{\bf k} = (\psi_{1, {\bf k}},\psi^{\dagger}_{1, -{\bf k}},\psi_{2, {\bf k}},\psi^{\dagger}_{2, -{\bf k}} )^T$ and write down the Hamiltonian for ${\bf \Psi}_{\bf k}$. Then, we see that ${\bf \Psi}^{\dagger}_{\bf k}{\bf \Psi}_{\bf k}=2$ (number constraint) and ${\bf \Psi}^{\dagger}_{\bf k}\sigma^{0}\otimes\sigma^{3}{\bf \Psi}_{\bf k}=0$ (particle-hole symmetry) for all ${\bf k}$. In the presence of translational symmetry, phase transitions between different $Z_2$ spin liquids happen when there is a level crossing at the high-symmetry points of the Brillouin zone\cite{Kou2009}. There are four such points: $(0,0), (0,\pi),(\pi,0), (\pi,\pi)$. At these points, fermion occupation number $n_{\bf k}\equiv\Psi_{\bf k}^\dagger\sigma_3\otimes\sigma_3\Psi_{\bf k}$ can take different values at these time-reversal-invariant momenta (TRIM) while $n_{\bf k} = 2$ for all other possible ${\bf k}$ in the Brillouin zone~\cite{Kou2008}. Different fillings at TRIM are the origin of the crystal momenta of Z2 spin liquids. The ground states can be obtained by filling all the states below the chemical potential $\mu =0$ (half-filled). Hence the ground state carries the crystal momentum depending on the filling $n_{\bf k}$ at these high-symmetry points.

Naively, there are four ground states for a Z2 spin liquid on the torus, and the four ground states can be labelled by the boundary conditions along the two directions of the torus  (the four ground states are generated by imposing `periodic'$\times$`periodic', `periodic'$\times$`anti-periodic', `anti-periodic'$\times$`periodic', `anti-periodic'$\times$`anti-periodic' boundary conditions along the two directions of the torus to the fermions). As the total number of the fermions in the spin liquid should be even and thus we require $\sum_{{\bf k} \in BZ} n_{{\bf k}} \in 2{\mathbb Z}$ to be a physical state. It is not difficult to see that the requirement $\sum_{{\bf k} \in BZ} n_{\bf k} \in 2{\mathbb Z}$ should reduce to $\sum_{{\bf k} \in TRIM} n_{\bf k} \in 2{\mathbb Z}$  because $n_{\bf k} = 2$ for all ${\bf k}\in BZ \setminus  TRIM$~\cite{Kou2008}. However, not all of ${\bf k} \in TRIM$ is allowed for the fermion. For example, on the odd-by-even lattice (with the size of the system $N_{x} \times N_{y}$ such that $N_{x} \in 2{\mathbb Z}+1$ and $N_{y} \in 2{\mathbb Z}$) with the periodic boundary conditions along the two directions of the torus, the allowed momenta $(k_{x}, k_{y})$ for the fermions can take the values of $(\frac{2\pi n_{x}}{N_{x}}, \frac{2\pi n_{y}}{N_{y}})$ with $n_{x} = 0, 1, 2, \cdots N_{x}-1, N_{x}$ and $n_{y} = 0, 1, 2, \cdots N_{y}-1, N_{y}$. In this case, there is no allowed state for the fermions at ${\bf k} = (\pi, 0)$ and $(\pi, \pi)$, and thus the summation $\sum_{{\bf k} \in TRIM} n_{{\bf k}}$ should be modified to the summation $\sum_{{\bf k} \in TRIM^{*}} n_{{\bf k}}$ where $TRIM^{*}$ is the set of the {\it allowed} momenta among ${\bf k} \in TRIM$. On the other hand, if the boundary condition along ${\hat x}$-direction is changed to be anti-periodic while keeping the boundary condition along ${\hat y}$ - direction periodic, then the allowed momenta $(k_{x}, k_{y})$ is changed to $k_{x} = \frac{2\pi}{N_{x}} (n_{x}+\frac{1}{2}), n_{x} = 0,1,2, \cdots N_{x}$ with $k_{y}$ as before. For this case, $TRIM^{*}$ is identical to $TRIM$ i.e., $TRIM^{*}= \{(0,0), (\pi, 0), (0,\pi), (\pi,\pi) \}$.

In the given lattice with the specificed boundary condition, we can compute the sum $\sum_{{\bf k} \in TRIM^{*}} n_{{\bf k}}$ for a ${\mathbb Z}_{2}$ spin liquid. If $\sum_{{\bf k} \in TRIM^{*}} n_{{\bf k}}\in 2{\mathbb Z}+1$, then the state labelled by the boundary condition is not a physical state and the state should not be considered as one of the ground states. Thus, the number of the ground states on the lattice can be smaller than the naive expectation $4$ on torus. When $\sum_{{\bf k} \in TRIM^{*}} n_{{\bf k}}\in 2{\mathbb Z}$, then the crystal momenta for the state is given by $\sum_{{\bf k} \in TRIM^{*}} {\bf k}n_{{\bf k}}$. Hence, if a Z2 spin liquid is given, we can compute the number of ground states on the four lattices (even-by-even, even-by-odd, odd-by-even, and odd-by-odd lattices) and label the ground states of the four lattice with the crystal momenta. For example, there are only two ground states for Wen's plaquette model on the even-by-odd lattices, and the two states carry the crystal momenta $(0,0)$ obtained by computing $\sum_{{\bf k} \in TRIM^{*}} {\bf k}n_{{\bf k}}$ for the two ground states. We can write this compactly as follows (we denote ``e$\times$o'' lattice as even-by-odd lattice) \newline${}$ \newline(ex) Wen's plaquette model\newline
${}\qquad$ e$\times$o lattice: $(0,0),(0,0)$ \newline ${}$

Following the above discussion, we find the crystal momenta of the ground states for the three Z2 spin liquids in the previous section (here we again denote ``e$\times$e lattce'' as even-by-even lattice, and ``e$\times$o lattice'' as even-by-odd lattices, etc.)  \newline{}\newline
(1) Wen's plaquette model, and ${\hat n} = {\hat x} \pm {\hat y}$\newline
${}\qquad$ e$\times$e lattice: $(\pi,\pi),(0,0),(0,0),(0,0)$\newline
${}\qquad$ e$\times$o lattice: $(0,0),(0,0)$\newline
${}\qquad$ o$\times$e lattice: $(0,0),(0,0)$\newline
${}\qquad$ o$\times$o lattice: $(0,0),(0,0)$\newline
(2) ${\hat n} = {\hat x}$\newline
${}\qquad$ e$\times$e lattice: $(0,\pi),(0,0),(0,0),(0,0)$\newline
${}\qquad$ e$\times$o lattice: $(0,0),(0,0)$\newline
${}\qquad$ o$\times$e lattice: $(0,\pi),(0,0),(0,0),(0,0)$\newline
${}\qquad$ o$\times$o lattice: $(0,0),(0,0)$\newline
(3) ${\hat n} = {\hat y}$\newline
${}\qquad$ e$\times$e lattice: $(\pi,0),(0,0),(0,0),(0,0)$\newline
${}\qquad$ e$\times$o lattice: $(\pi,0),(0,0),(0,0),(0,0)$\newline
${}\qquad$ o$\times$e lattice: $(0,0),(0,0)$\newline
${}\qquad$ o$\times$o lattice: $(0,0),(0,0)$\newline {}

We will see that this complicated pattern of the crystal momentum can be reproduced by the effective $BF$ theory in the next subsection. While deriving the above result, we implicitly assumed that the Hamiltonian $Q(\{f \}) = \pm 1$ for the trivial chain not to have any potential structure in it. While the Hamiltonian $Q$ here is artificial, we need to close the gap at the high-symmetry points of the Brillouin zone to change the crystal momenta. Hence, the pattern of the crystal momenta should be properties of a phase, not of a point in the phase diagram.

\subsection{Classification of $BF$ theory on the square lattice with translational symmetry}

To classify $BF$ theory, we begin with the lesson from the previous studies\cite{Wen2003,Kou2008} that we will utilize in this section. The first lesson is that encoding the transformation on the gauge theory is not 'gauge' degree of freedom, but it is a physical symmetry. For example in the toric code or Wen's plaquette model, we exchange the electric excitation and the magnetic excitation under the translation. This is different from $SL(2;{\mathbb Z})$ gauge symmetry for $BF$ theory. The second lesson is that we can obtain the complicated patterns for the crystal momenta and degeneracies by implementing the non-trivial transformations. Hence, the strategy is simple: finding all the possible physical transformations for $BF$ field theory. In the following discussion, we will systematically search and implement the transformations for $BF$ theory.

We begin with the Coulomb gauge $a_{0}=b_{0}=0$ for the gauge fields $a_{\mu}$ and $b_{\mu}$ in Eq.\eqref{BF}. Then, this reduces Eq.\eqref{BF} into the following form (without sources)
\beq
L = \frac{1}{2\pi} {\bf v}^{T} \cdot {\bf J} \cdot  \partial_{0}{\bf v}
\label{Vector}
\eeq
where ${\bf v}^{T} = (a_{x}, a_{y}, b_{x},b_{y})$ and
\begin{align}
{\bf J}= \left[
\begin{array}{cccc}
0 & 0 & 0 & +1 \\
0 & 0 & -1 & 0 \\
0 & +1 & 0 & 0 \\
-1 & 0 & 0 & 0
\end{array}
\right]
\end{align}
To be invariant under a unit lattice translation $t_{x}$ in $x$- axis (and similar for $t_{y}$ in $y$-axis), we require
\beq
L = \frac{1}{2\pi} {\bf v}^{T} \cdot {\bf J} \cdot  \partial_{0}{\bf v} =  \frac{1}{2\pi} (t_{x}\left[ {\bf v}^{T}\right]) \cdot {\bf J} \cdot  \partial_{0}(t_{x}\left[ {\bf v}^{T}\right]).
\label{requirement}
\eeq
Hence, we look for solutions $t_{x}$ and $t_{y}$ satisfying Eq.\eqref{requirement}. The obvious possibility for $t_{x}$ and $t_{y}$ is to consider the linear transformations acting on ${\bf v}$, i.e., we associate $(t_{x}, t_{y})$ with the matrices $(O_{x}, O_{y})$ and the vectors $({\bf u}_{x}, {\bf u}_{y})$ such as
\beq
t_{x, y}: {\bf v} \rightarrow O_{x, y}{\bf v} + {\bf u}_{x, y}.
\eeq
We plug this into Eq.\eqref{requirement} and find a restriction on $O_{x,y}$
\beq
O_{x,y}^{T} J O_{x,y} = J,
\label{constraint}
\eeq
The restriction for ${\bf u}_{x,y}$ shows up when we take the microscopic picture into consideration. Before proceeding further, we notice that $O_{x,y}$ is orthogonal, i.e., det$(O_{x,y})$=det$(O^{T}_{x,y}) = \pm 1$ as $J$ is invertible. This alone cannot fix the form of $O_{x,y}$. We also note that $O_{x,y}^{2} =1$. This is because the size of the smallest even-by-even lattice is $2\times 2$, and we expect the degeneracies on the even-by-even lattice are trivially $4$ i.e., we force the gauge field ${\bf v}$ to return to itself. Then, it is easy to check that there are four solutions for $O_{x,y}$ satisfying Eq.\eqref{constraint} (Note that the overall sign of the $O_{x,y}$ is irrelevant as the sign of ${\mathbb Z}_{2}$ theory is not important). The first physical solution is the trivial ${\bf I} =$diag$(1,1,1,1)$. The second physical solution is `twist' ${\bf O}^{t}$
\begin{align}
{\bf O^{t}}= \left[
\begin{array}{cccc}
0 & 0 & 1 & 0 \\
0 & 0 & 0 & 1 \\
1 & 0 & 0 & 0 \\
0 & 1 & 0 & 0
\end{array}
\right]
\end{align}
which exchanges $a_{i}$ and $b_{i}$, i.e., $a_{i} \rightarrow b_{i}$ and $b_{j} \rightarrow a_{j}$ under a unit lattice translation. The third possible (but not physical) solution ${\bf Q}$ is
\begin{align}
{\bf Q}= \left[
\begin{array}{cccc}
0 & 1 & 0 & 0 \\
1 & 0 & 0 & 0 \\
0 & 0 & 0 & -1 \\
0 & 0 & -1 & 0
\end{array}
\right]
\end{align}
which acts as $(a_{x}, b_{x}) \rightarrow (a_{y}, -b_{y})$ and $(a_{y}, b_{y})\rightarrow (a_{x}, -b_{x})$. The fourth non-physical solution ${\bf P}$ is
\begin{align}
{\bf P}= \left[
\begin{array}{cccc}
1 & 0 & 0 & 0 \\
0 & -1 & 0 & 0 \\
0 & 0 & -1 & 0 \\
0 & 0 & 0 & 1
\end{array}
\right]
\end{align}
which flips the sign of $a_{y}$ and $b_{x}$ spontaneously. A physical argument for excluding ${\bf P}$ and ${\bf Q}$ comes from the coupling of $BF$ fields to the ${\mathbb Z}_{2}$ charge field $z$ and the ${\mathbb Z}{2}$ vortex field $v$. We have the full Lagrangian $L = L_{BF} + L_{coup}$ where $L_{BF}$ is Eq.\eqref{Vector} and (in the Coulomb gague $a_{0}=b_{0}=0$)
\beq
L_{coup} = |(\partial_{j} - ia_{j})z|^{2} + |(\partial_{j} - ib_{j})v|^{2},
\eeq
Now, we translate the system by one unit lattice along $x$-axis which acts as $t_{x}$ on the fields.  Here, $t_{x}\left[ \partial_{\mu}\right] = \partial_{\mu}$ and $t_{x}\left[L_{BF}\right] = L_{BF}$ , and we are required to satisfy the equation $L_{coup} = t_{x}\left[L_{coup}\right]$ where
\beq
t_{x}\left[L_{coup}\right] =   |(\partial_{j} - it_{x}\left[a_{j}\right])t_{x}\left[z\right]|^{2} + |(\partial_{j} - it_{x}\left[b_{j}\right])t_{x}\left[v\right]|^{2}.
\eeq
For $t_{x} = {\bf I}$, we have $t_{x}: z \rightarrow z$ and $t_{x}: v \rightarrow v$. For $t_{x} = {\bf O}^{t}$, we have $t_{x}: z \rightarrow v$ and $t_{x}: v \rightarrow z$. However, for ${\bf P}$ and ${\bf Q}$, there is no way to make the equality $t_{x}\left[L_{coup}\right] = L_{coup}$ from $t_{x}\left[z\right]$ and $t_{x}\left[v\right]$, hence we exclude them as the physical transformation of $BF$ theory which is consistent with the Z2 spin liquids in the previous section.

Now, we move on to the constant vector ${\bf u}_{x,y}$. As the vectors ${\bf v}$ are ${\mathbb Z}_{2}$ gauge theory in the microscopic picture, so should be ${\bf u}_{x,y}$. More specifically, $\left[{\bf u}_{x,y}\right]_{a} \in \{0, \pi \}$ mod $2\pi$. However, not all of ${\bf u}_{x,y}$ pattern are physically independent. There are $\pi$ flux or no flux in ${\mathbb Z}_{2}$ gauge theory for $a_{i}$ and $b_{j}$, and this reduces many ${\bf u}_{x,y}$ patterns into only four of them. Those are the followings: $1)$ Both $a_{\mu}$ and $b_{\nu}$ have fluxes in the unit cell. $2)$ only $a_{\mu}$ contains fluxes in the unit cell. $3)$ only $b_{\nu}$ contains fluxes in the unit cell. $4)$ No fluxes are present. For $1)$, we have ${\bf u}_{x}^{T} = (0, \pi,0,0)$ and ${\bf u}_{y}^{T} = (0, 0,\pi,0)$. For $2)$, we have ${\bf u}_{x}^{T} = (0, \pi,0,0)$ and ${\bf u}_{y}^{T} = (0,0,0,0)$. For $3)$, we have ${\bf u}_{x}^{T} = (0, 0,0,0)$ and ${\bf u}_{y}^{T} = (0,0,\pi,0)$. For $4)$, we have ${\bf u}_{x}^{T} = {\bf u}_{y}^{T} = (0, 0,0,0)$.

We write $(t_{x}, t_{y})$  as $\{({\bf O}_{x}, {\bf u}_{x}), ({\bf O}_{y}, {\bf u}_{y})\}$. We denote twist matrix simply as ${\bf O}^{t}$ and trivial matrix as ${\bf I}$, and we also denote the nontrivial flux vector as ${\bf u}$ and no flux as ${\bf 0}$. For example, we can have $BF$ theory with the translational invaraint property $\{({\bf O}^{t}, {\bf 0}), ({\bf I}, {\bf u}_{y})\}$ satisfies the following relations: $t_{x}: (a_{i}, b_{j}) \rightarrow (b_{i}, a_{j})$ and $t_{y}: (a_{i}, b_{j}) \rightarrow (a_{i}, b_{j} + \pi \delta_{x, j})$.

A simple combinatorial computation then concludes that this classification gives $2 \times 2 \times 4  = 16$ classes for $BF$ theory consistent with the underlying ${\mathbb Z}_{2}$ gauge theory on the square. Among $16$ classes, six classes are not of interest. To see this, we look at the commutation relation for $t_{x}$ and $t_{y}$
\beq
\left[ t_{x}, t_{y} \right] \neq 0
\eeq
seemingly implying that those states are not physical. But this is not true; for example, the magnetic translations are not commuting each other in the quantum Hall states. Still, there is no allowed ground state\cite{Kou2008} in the odd-by-odd lattices if $\left[ t_{x}, t_{y} \right] \neq 0$. As all of our targeting states have $2$ or $4$ states, we exclude the $BF$ theory of $\left[ t_{x}, t_{y} \right] \neq 0$.

\subsection{Crystal momenta, degeneracies, and ${\mathbb Z}_{2}$ indices}
Upon obtaining the transformations for the relevant $BF$ theory, we compute the crystal momenta of the ground states on even-by-even, odd-by-even, even-by-odd, and odd-by-odd lattices. We follow the straightforward calculation in the reference [\onlinecite{Kou2008}] to get the crystal momenta.

Before presenting a series of  results from the computation, we exhibit two examples showing how we get the crystal momenta on lattices.
{}\newline\newline
{\it Example 1}. $\{({\bf I}, {\bf u}_{x}), ({\bf I}, {\bf 0})\}$ \newline
We have the transformations under the translations as $t_{x}: (a_{i}, b_{j}) \rightarrow (a_{i}+\pi\delta_{y,i}, b_{j})$ and $t_{y}:(a_{i}, b_{j}) \rightarrow (a_{i}, b_{j})$. We denote the zero mode of $BF$ fields $(a_{i}, b_{j})$ as $(\theta_{i}, \phi_{j})$. Then, the $BF$ theory relates $\theta_{x}$ ($\phi_{x}$) to $\phi_{y}$ ($\theta_{y}$) as the canonical conjugate pairs. Explicitly, we have $\left[\theta_{x}, \phi_{y}\right] = i\pi$ and $\left[\phi_{x},\theta_{y}\right]=i\pi$. Then, we have four well-defined ground states $(|1>, |2>,|3>,|4>)$ such as $|2>=e^{-i\theta_{x}}|1>, |3>=e^{-i\phi_{x}}|1>, |4>=e^{-i\theta_{x}}e^{-i\phi_{x}}|1>$ where $|1>$ transforms trivially under any transformations of $t_{x,y}$. Then, we have $t_{x}: (|1>, |2>,|3>,|4>) \rightarrow (|3>, |4>,|1>,|2>)$ and $t_{y}: (|1>, |2>,|3>,|4>) \rightarrow (|1>, |2>,|3>,|4>)$. Hence, $[\ t_{x}, t_{y} ]\ = 0$ and $t_{x}$ and $t_{y}$ can be simultaneously diagonalized. We conclude that the two states of the form $\frac{1}{\sqrt{2}}(|1>-|3>)$ and $\frac{1}{\sqrt{2}}(|2>-|4>)$ carry the crystal momentum $(\pi,0)$, and other two states carry the crystal momentum $(0,0)$ for any lattice. This gives the crystal momentum spectrums $(\pi,0), (\pi,0), (0,0), (0,0)$ on even-by-even, even-by-odd, odd-by-even, and odd-by-odd lattices, and there is no matched Z2 spin liquids of interest.
{}\newline\newline
{\it Example 2}. $\{({\bf O}^{t}, {\bf 0}), ({\bf I}, {\bf 0})\}$ \newline
We have the transformation law under the translations $t_{x}: (a_{i}, b_{j}) \rightarrow (b_{i}, a_{j})$ and $t_{y}:(a_{i}, b{j}) \rightarrow (a_{i}, b_{j})$. We begin with the even-by-even lattice. Then, we can expand $BF$ theory and obtain the zero modes $\theta_{x}$ ($\phi_{x}$), canonical conjugate to $\phi_{y}$ ($\theta_{y}$). Then as before, we have four well-defined ground states $(|1>, |2>,|3>,|4>)$ such as $|2>=e^{-i\theta_{x}}|1>, |3>=e^{-i\phi_{x}}|1>, |4>=e^{-i\theta_{x}}e^{-i\phi_{x}}|1>$ where $|1>$ transforms trivially under any transformations of $t_{x,y}$. Then, we have $t_{x}: (|1>, |2>,|3>,|4>) \rightarrow (|1>, |3>,|2>,|4>)$ and $t_{y}: (|1>, |2>,|3>,|4>) \rightarrow (|1>, |2>,|3>,|4>)$. Hence, $[\ t_{x}, t_{y} ]\ = 0$ and $t_{x}$ and $t_{y}$ can be simultaneously diagonalized. We conclude that one state of the form $\frac{1}{\sqrt{2}}(|2>-|3>)$ carries the crystal momentum $(\pi,0)$, and the other three states carry the crystal momentum $(0,0)$ for even-by-even lattice. This gives the spectrum of the crystal momenta $(\pi,0), (0,0), (0,0), (0,0)$ on even-by-even lattice. Now, we move to the even-by-odd lattice. In this case, we can expand the zero mode as well as in even-by-even lattice because $t_{y}$ is trivial, and this gives the crystal momentum spectrum $(\pi,0), (0,0), (0,0), (0,0)$ on even-by-odd lattice. It turns out to be different for odd-by-even lattice as the non-trivial boundary condition $T_{x}:(a_{i}, b_{j}) \rightarrow (b_{i}, a_{j})$ where $T_{x} = (t_{x})^{L_{x}}$, and we cannot expand $BF$ theory to obtain the zero mode. The resolution for this is to double the odd-by-even lattice along $x$- axis to get even-by-even lattice. Then, we can now expand $BF$ fields to obtain the zero modes $(|1>, |2>,|3>,|4>)$ such as $|2>=e^{-i\theta_{x}}|1>, |3>=e^{-i\phi_{x}}|1>, |4>=e^{-i\theta_{x}}e^{-i\phi_{x}}|1>$ with the boundary condition $T_{x}: (\theta_{x}, \phi_{x})\rightarrow (\phi_{x},\theta_{x})$. The boundary condition implies that the originally four independent states bind to each other to form only two independet states $|1>, |2>$. Furthermore, $t_{x}: (|1>, |2>) \rightarrow (|1>,|2>)$, and thus the crystal momentum spectrum on the odd-by-even lattice is $(0,0), (0,0)$. The similar consideration on odd-by-odd lattice gives the crystal momentum spectrum as $(0,0), (0,0)$. These crystal momentum spectrums match the spin liquids of ${\hat n}={\hat y}$ in the previous section. \newline\newline

This calculation can be also done for other translational symmetric $BF$ theory, and now we find that
\newline\newline
(1) $(\{({\bf O}^{t}, {\bf 0}), ({\bf O}^{t}, {\bf 0})\})$ is Wen's model or ${\hat n}= {\hat x}\pm {\hat y}$\newline
(2) $(\{({\bf O}^{t}, {\bf 0}), ({\bf I}, {\bf 0})\})$ is ${\hat n} = {\hat y}$ Z2 spin liquid \newline
(2) $(\{({\bf I}, {\bf 0}), ({\bf O}^{t}, {\bf 0})\})$ is ${\hat n} = {\hat x}$ Z2 spin liquid \newline

It is not difficult to see that these $BF$ theories have the same edge spectrum as the micrscopic consideration in the previous section. The spectrum of the crystal momenta and degeneracies for other $BF$ theories can be found in the appendix.

Before finishing this section, we comment on the connection to the previous study\cite{Kou2009}. The pattern of the crystal momenta and degeneracies of Z2 spin liquids are, in fact, fixed by the fermion parity at the high symmetry points in the Brillouin zone, and this fact can be used to classify Z2 spin liquids. We can find that Wen's model and ${\hat n} = {\hat x} \pm {\hat y}$ belong to $[0110]$ and $[1001]$ classes in the classification of Z2 spin liquids in the index system of the previous study (${\hat n} = {\hat x}$ belongs to $[1100]$ and $[0011]$ classes, and ${\hat n} = {\hat y}$ belongs to $[1010]$ and $[0101]$ classes.) Though it can be shown that these Z2 spin liquids with the gapless edge belong to the particular class, it is not clear if this index system implies the gapless edge states between Z2 spin liquids and vacuum. It would be an interesting future direction to study if the indices imply the presence of the gapless edge state.

\section{Generalizations and Conclusions}

Can we generalize our reasoning for the gapless edge states of Z2 spin liquids on the square lattice to other lattices such as the triangular lattices? More interestingly, can we generalize it to higher dimensions? It is not difficult to note that the answers to these questions are `yes', at least in the mean-field Hamiltonian. The nature of the edge states will depend on the direction of stacking one-dimensional Majorana fermion chains for $f_{i,u}$ and $f_{i,d}$. Hence, we can apply this to any two-dimensional lattice to construct Z2 spin liquids with the gapless edge states. We can also generalize to three-dimensional spin liquids to write down the mean field Hamiltonian, but it is better to have an exactly solvable model that exhibits the gapless surface states for the cubic lattice. The model is the direct generalization of Wen's plaquette model on the square lattice, and it involves six Majorana fermions per site, i.e., this is a Hamiltonian for a spin-$3/2$ system\cite{Levin2003}. We start with the ${\mathbb Z}_{2}$ gauge theory on the cubic lattice
\beq
H = -g \sum_{P} F_{P} = -g \sum_{P} \prod_{<ij>\in P} U_{ij},
\label{Wen2}
\eeq
where $P$ denotes the plaquette of the cubic lattice. Here, $U_{ij} = \pm 1$ is the ${\mathbb Z}_{2}$ gauge on the link $<ij>$, and thus $ \prod_{<ij>\in P} U_{ij} = F_{P}$ is the field strength on the plaquette $P$. The particular (but general) representation of this gauge theory is $U_{i,i+{\hat x}} = i \lambda_{i,x}\chi_{i+{\hat x},x}$, $U_{i,i+{\hat x}} = i \lambda_{i,y}\chi_{i+{\hat y},y}$, $U_{i,i+{\hat z}} = i \lambda_{i,z}\chi_{i+{\hat z},z}$ with the translational symmetry. Hence, we have the six Majorana fermions $(\lambda_{i,x},\lambda_{i,y}, \lambda_{i,z}, \chi_{i,x}, \chi_{i,y},\chi_{i,z})$ per site, and the dimension of the Hilbert space per site is $2^{6/2}= 8$ states per site. Again we need to halve this Hilbert space to represent a spin-3/2 system,, i.e., the Hilbert space per site should be $4$. This is done by taking the ${\mathbb Z}_{2}$ redundancy into account: Eq.\eqref{Wen2} is invariant under $U_{ij} \rightarrow s_{i}U_{ij}s_{j}$, $s_{i} = \pm1$.

In the three-dimensional model, we see that all the Majorana fermions in the bulk are paired. On the boundary, however we have one dangling Majorana fermion per unit cell on the boundary. Following the reasoning in the two-dimensional system, the flat bands on the boundary will start to disperse to form the gapless surface states when a small perturbation away from this exactly solvable limit induces
\beq
H_{edge} =  i \sum_{i}(t_{x} \eta_{i}\eta_{i+{\hat x}} + t_{y} \eta_{i}\eta_{i+{\hat y}}).
\eeq
This Hamiltonian has the gapless spectrum $E({\bf k}) = \sqrt{t^{2}_{x}\sin^{2}(k_{x})+t^{2}_{y}\sin^{2}(k_{y})}$. To gap out the spectrum, we need to double the unit cell\cite{Mong2012,Ringel2011} which is prohibited by the translational symmetry.

As before, we can deform the exactly solvable Hamiltonian to obtain a series of anisotropic Z2 spin liquids with the gapless surface states. How many different classes of the Z2A spin liquids with translational symmetry in the cubic lattice are there (if two Z2 spin liquids have the same crystal momentums and degeneracies, and the surface states, we define them as the same class)? We repeat the argument in the two-dimensional case and construct 3D mean-field (ideal) Hamiltonians of $Z_2$ spin liquids from the one-dimensional Majorana fermion chains. By checking the crystal momenta and the degeneracies in the three-dimensional lattices, we found that there are $14$ classes which have the gapless surface states protected by the translational symmetry.  The effective theory for these $Z_2$ spin liquids should be (3+1)- dimensional $BF$ theory\cite{Hansson2004}
\beq
L = \frac{1}{2\pi}b_{\mu\nu}\partial_{\lambda} a_{\rho} - \frac{1}{2} \Sigma^{\mu\nu}b_{\mu\nu} - a_{\mu}j^{\mu},
\eeq
which is known to have the gapless Dirac-like surface spectrum\cite{Cho2011}. It would be an interesting future research direction to show how the translational invariance stabilizes the gapless surface states of (3+1)-dimensional $BF$ theory .

In summary, we considered translationally-symmetric Z2 spin liquids which have gapless edge/surface states. The edge/surface states are constructed out of the dangling Majorana fermions on the boundary. We clarified the conditions for the gapless edge states for Z2 spin liquids and constructed the effective $BF$ theory reflecting the underlying translational symmetry of the lattice. The general reasoning applies to the three-dimensional lattices, and we demonstrated that there is a gapless surface state.

\acknowledgments
The authors thank Z.C Gu, E.A Kim, K. Shtengel, F. Burnell, E.G Moon, and C. Laumann for helpful discussions. Funding support for this work was provided by the KITP Graduate Fellowship and FENA (G.Y.C); Office of BES, Materials Sciences Division of the U.S. DOE under contract No. DE-AC02-05CH1123 (Y.M.L.) and NSF DMR-0804413 (J.E.M.).\\

\appendix


\section*{Crystal Momenta of $BF$ theory}
We list the crystal momenta of each theory (we denote ``e$\times$e lattce'' as even-by-even lattice, and ``e$\times$o lattice'' as even-by-odd lattices, etc)
\newline\newline (1) $(\{({\bf I}, {\bf 0}), ({\bf I}, {\bf 0})\})$ is $\left[0000\right]$ Z2 spin liquid\newline
 (2) $(\{({\bf I}, {\bf u}_{x}), ({\bf I}, {\bf 0})\})$ \newline
${}\qquad$ e$\times$e lattice: $(\pi,0),(\pi,0),(0,0),(0,0)$\newline
${}\qquad$ e$\times$o lattice: $(\pi,0),(\pi,0),(0,0),(0,0)$\newline
${}\qquad$ o$\times$e lattice: $(\pi,0),(\pi,0),(0,0),(0,0)$\newline
${}\qquad$ o$\times$o lattice: $(\pi,0),(\pi,0),(0,0),(0,0)$\newline
 (3) $(\{({\bf I}, {\bf 0}), ({\bf I}, {\bf u}_{y})\})$ \newline
${}\qquad$ e$\times$e lattice: $(0,\pi),(0,\pi),(0,0),(0,0)$\newline
${}\qquad$ e$\times$o lattice: $(0,\pi),(0,\pi),(0,0),(0,0)$\newline
${}\qquad$ o$\times$e lattice: $(0,\pi),(0,\pi),(0,0),(0,0)$\newline
${}\qquad$ o$\times$o lattice: $(0,\pi),(0,\pi),(0,0),(0,0)$\newline
 (4) $(\{({\bf I}, {\bf u}_{x}), ({\bf I}, {\bf u}_{y})\})$ is $\left[1111\right]$ Z2 spin liquid\newline
(5) $(\{({\bf O}^{t}, {\bf 0}), ({\bf O}^{t}, {\bf 0})\})$ is $\left[0110\right],\left[1001\right]$ Z2 spin liquid\newline
 (6) $(\{({\bf O}^{t}, {\bf u}_{x}), ({\bf O}^{t}, {\bf 0})\})$ ($\left[t_{x},t_{y}\right]\neq 0$) \newline
 (7) $(\{({\bf O}^{t}, {\bf 0}), ({\bf O}^{t}, {\bf u}_{y})\})$ ($\left[t_{x},t_{y}\right]\neq 0$) \newline
 (8) $(\{({\bf O}^{t}, {\bf u}_{x}), ({\bf O}^{t}, {\bf u}_{y})\})$\newline
${}\qquad$ e$\times$e lattice: $(\pi,\pi),(\pi,0),(0,\pi),(0,0)$\newline
${}\qquad$ e$\times$o lattice: $(0,0),(0,0)$\newline
${}\qquad$ o$\times$e lattice: $(0,0),(0,0)$\newline
${}\qquad$ o$\times$o lattice: $(0,0),(0,0)$\newline
(9) $(\{({\bf O}^{t}, {\bf 0}), ({\bf I}, {\bf 0})\})$ is $\left[0101\right], \left[1010\right]$ Z2 spin liquid \newline
(10) $(\{({\bf O}^{t}, {\bf 0}), ({\bf I}, {\bf u}_{y})\})$ ($\left[t_{x},t_{y}\right]\neq 0$)\newline
(11) $(\{({\bf O}^{t}, {\bf u}_{x}), ({\bf I}, {\bf 0})\})$ \newline
${}\qquad$ e$\times$e lattice: $(\pi,0),(\pi,0),(0,0),(0,0)$\newline
${}\qquad$ e$\times$o lattice: $(\pi,0),(\pi,0),(0,0)(0,0)$\newline
${}\qquad$ o$\times$e lattice: $(0,0),(0,0)$\newline
${}\qquad$ o$\times$o lattice: $(0,0),(0,0)$ \newline
(12) $(\{({\bf O}^{t}, {\bf u}_{x}), ({\bf I}, {\bf u}_{y})\})$ ($\left[t_{x},t_{y}\right]\neq 0$)\newline
(13) $(\{({\bf I}, {\bf 0}), ({\bf O}^{t}, {\bf 0})\})$ is $\left[1100\right], \left[0011\right]$ Z2 spin liquid \newline
(14) $(\{({\bf I}, {\bf u}_{x}), ({\bf O}^{t}, {\bf 0})\})$ ($\left[t_{x},t_{y}\right]\neq 0$)\newline
(15) $(\{({\bf I}, {\bf 0}), ({\bf O}^{t}, {\bf u}_{y})\})$ \newline
${}\qquad$ e$\times$e lattice: $(0,\pi),(0,\pi),(0,0),(0,0)$\newline
${}\qquad$ e$\times$o lattice: $(0,0)(0,0)$\newline
${}\qquad$ o$\times$e lattice: $(0,\pi),(0,\pi),(0,0),(0,0)$\newline
${}\qquad$ o$\times$o lattice: $(0,0),(0,0)$ \newline
(16) $(\{({\bf I}, {\bf u}_{x}), ({\bf O}^{t}, {\bf u}_{y})\})$ ($\left[t_{x},t_{y}\right]\neq 0$)\newline

We looked at the patterns of the crystal momenta of the Z2 spin liquids\cite{Kou2009} and found four ${\mathbb Z}_{2}$ indices of Z2 spin liquids which match these patterns. Note that the classes $(5)-(16)$ of $BF$ theories have the gapless edge state.


\end{document}